# Compositional memory systems for multimedia communicating tasks

A.M. Molnos[(*)(**)]  M.J.M. Heijligers[(**)]  S.D. Cotofana[(*)]  J.T.J. van Eijndhoven[(**)]

[(*)] Delft University of Technology
Mekelweg 4, Delft, The Netherlands
molnos@natlab.research.philips.com

[(**)] Philips Research Laboratories
Prof. Holstlaan 4, 5656 AA
Eindhoven, The Netherlands

**Abstract**

*Conventional cache models are not suited for real-time parallel processing because tasks may flush each other's data out of the cache in an unpredictable manner. In this way the system is not compositional so the overall performance is difficult to predict and the integration of new tasks expensive. This paper proposes a new method that imposes compositionality to the system's performance and makes different memory hierarchy optimizations possible for multimedia communicating tasks when running on embedded multiprocessor architectures. The method is based on a cache allocation strategy that assigns sets of the unified cache exclusively to tasks and to the communication buffers. We also analytically formulate the problem and describe a method to compute the cache partitioning ratio for optimizing the throughput and the consumed power. When applied to a multiprocessor with memory hierarchy our technique delivers also performance gain. Compared to the shared cache case, for an application consisting of two jpeg decoders and one edge detection algorithm 5 times less misses are experienced and for an mpeg2 decoder 6.5 times less misses are experienced.*

## 1. Introduction

The system's high performance and predictability are the required characteristics for state-of-the-art embedded multimedia applications. To meet their performance requirements parallel processing of data on multi-processor architectures is needed. The low cost demands of this application domain make the use of general purpose architectures with several GHz clock frequencies impossible.

The applications' increase in size and complexity demand hardware platforms with large guaranteed memory bandwidth. The increase in bandwidth to off-chip memories is not growing as fast as the increase in speed of computation units. A possible approach to cope with this processor-memory gap is to use memory hierarchies (caches) [3]. For embedded systems caches induce undesired unpredictability because tasks can influence each other's performance by flushing each other's data out of the cache. In this way the system is not compositional so the overall performance cannot be predicted based on the system's parts performance and changing of components has design implications to the overall system causing longer time-to-market with economical implications.

In the real-time domain one could think of replacing caches with scratch pad memories, but in practice removing caches is not a good choice because they provide flexibility. They don't need to be resized or modified if the application changes. Detailed analysis regarding the memory requirements of multimedia applications' code and data at every change of standards or features takes much effort and negatively affects the time-to-market.

The importance of managing caches in general was identified already in the single processor with multiple threads domain. Several cache partitioning and management techniques are proposed for time-sharing environment [10], instruction controlled, compile time configured data cache [7], first level compiler-based partitioned cache [6]. However, for the multiprocessor domain there is a lack of cache management methods.

This article presents a new cache management technique applicable for multimedia applications when running on multiprocessor architectures that achieve systems' performance compositionality. The method allows tuning the memory system according to different optimization criteria. Our proposal is based on the observation that the following holds true: in a multiprocessor system, the levels of unified cache shared between processors are the most affected by the inter-task conflicts. Thus we propose to achieve compositionality by exclusively assigning sets of the last level of cache to tasks and to the communication buffers. An accurate analytical formulation and a practical approximation to find the cache partitioning ratio for throughput and power optimization are presented. The influence of task to processor assignment in this formulation is also discussed. This paper presents a continuation of [5] where the case of non-communicating multimedia applications is discussed.

To evaluate the performance implications of our method we consider two applications (two instances of jpeg decoder plus one canny edge detection algorithm and an mpeg2 decoder) running on a multiprocessor CAKE platform [9] with 4 VLIW processors and 512KB of L2 cache. The simulations indicate that when the proposed paradigm is applied to the L2 cache the system becomes compositional and also the number of L2 misses is diminished up to 5 times for the first applications and 6.5 times for the second application.

The outline of this paper is as follows. Section 2 overviews the state of the art in the field of cache management for multitasking environments. Section 3 discuss the proposed method to induce compositionality together with the optimization possibilities and an approach to determine the static cache partitioning ratio. Section 4 and 5 present the experimental framework and results. Finally, in Section 6 conclusions are drawn.

## 2. Related work

In the single processor domain different hardware ([10], [8], [4]), software ([6]) or mixed ([7]) cache management methods were proposed. In [10] the problem of cache efficiency for simultaneous threads in a time-sharing environment is tackled using dynamic column caching, in a best-effort way. Based on their number of misses tasks are dynamically "stealing" each other cache ways, such that the overall number of misses is improved. In [8] the problem of optimal allocation of cache between two competing processes that minimizes the overall miss rate is presented. In [4] a method to divide a cache into partitions for each real-time task and a larger partition called the shared pool for the non-real-time tasks is described. In [6] the cache is partitioned among tasks at compile and link time. In [7] a new data cache organization is proposed. The L1 direct mapped cache can be partitioned and configured compile time and controlled by specific cache instructions at run time, bringing considerable performance gain.

The main differences between our work and the previous research are that we tackle the case of applications running on platforms with unified set-associative cache shared between the processors. On such platforms, in our opinion without explicitly taking into account the inter-task communication, no compositionality can be achieved and performance cannot be guaranteed. None of the above approaches can be straightforwardly extended to our case. Column caching used [10] and [8] is based on ways allocation from every cache set to the tasks. This partitioning type severely restricts the granularity of cache allocation to the associativity of the cache. In [4] data access to the shared structures do not have predictable cache access. The method in [6] targets the L1 cache, requires a specific compiler and it is rather inflexible. The method in [7] is limited to L1, direct mapped data caches for uniprocessors.

## 3. Cache partitioning for system compositionality

The platform model for the proposed method is a homogeneous on-chip multiprocessor having high bandwidth communication network with the partitionable shared unified on-chip cache. On this platform parallel tasks that communicate through the memory system are executed. Resource contention for the shared buses and the shared levels of cache are generating unpredictable inter-task conflicts. The communication network has high bandwidth so the resource contention there is low. Since the levels of cache private to each processor are usually small and task switching rate in multimedia application is typically low enough, the first levels of cache can be considered private to each task. The method to induce compositionality to the system is by allocating to the tasks and the communication buffers exclusive parts of the level of cache that is shared through the processors.

The communication buffers addresses in cache are accessed by multiple tasks. If every task has a private cache part, assigning one buffer for example in the consumer's cache will allow the producer to unpredictably evict some consumer cached data. If all the buffers are allocated in a shared cache pool they will still unpredictably evict each other data, depending on their addresses and the timing of the productions and the consumptions. In order to have no cache interactions between tasks the communication buffers need their own exclusive cache partitions.

Let's now assume that the cache allocated for every communication buffer is smaller than its size and the production and consumption happens in parallel. If the producer is slow enough not to flush former produced data out of the cache the consumer will have a hit. Depending on the producing/consuming rates the number of misses varies. So, in order to have predictability, for the communication buffers one of the following should be made:

- ensure that all accesses are hits (except cold misses).
- ensure that all accesses miss in the cache.
- predict the number of misses at the buffer level - this is rather difficult because requires cycle accurate producer and consumer rates which are difficult to estimate and usually not constant in a multiprocessor environment.

### 3.1. Optimization opportunities

In a compositional environment the performance of one task is independent of the other tasks and this is achieved by cache partitioning. Different optimization criteria, statically prioritizing tasks and guaranteeing performance are possible by tuning the cache partitioning ratio.

An application consisting of $N$ tasks and $M$ communication buffers can be represented as a graph $G = (T, B)$, where $T = \{t_i\}_{(i=0,1...N-1)}$ is the set of nodes denoting the tasks and $B = \{b_{i,j}\}_{(i,j=0,1...N-1)}$ is the set of edges denoting inter-task communication. An edge $b_{i,j}$ is present in the application's graph if the task $t_i$ exchange data with the task $t_j$. We consider the case of homogeneous multiprocessors and the set of processors in the system is $P = \{p_k\}_{(k=0,1...R-1)}$.

Typically a multimedia application executes for an infinite time in a periodic manner. Every task $t_i$ has assigned the following functions: $c(t_i)$ expressing its allocated cache size and $e(t_i, c(t_i))$ expressing its execution time for one application period.

In a multiprocessor platform the execution time of every task $t_i$ is specified by a set $S_i = \{s_i^q\}_{(t_i \in T, q=0,1,...Q_i)}$ of time slices when $t_i$ executes on different processors. Depending on the cache size $c(t_i)$ the overall execution time of task $t_i$ becomes:

$$e(t_i, c(t_i)) = \sum_{j \in S_i} e(j, proc(j), c(t_i))$$

where $e(j, proc(j), c(t_i))$ is the execution time of slice $j$ and $proc(j)$ denotes the processor executing slice $j$.

For the multimedia applications a good optimization criteria is the execution throughput. The execution's throughput can be defined as the number of application's complete executions in a time unit. Let's denote with $E(p_k)$ the time processor $p_k$ is used for the completion of one application execution. The throughput is given by the $max_{(p_k \in P)}(E(p_k))$. Let $T_k \subseteq T$ be the subset of tasks that execute at least one of their slices on $p_k$, then the processor execution time becomes:

$$E(p_k) = \sum_{\substack{t_i \in T_k \\ j \in S_i \\ proc(j)=p_k}} e(j, p_k, c(t_i)) + t_{switch} + t_{idle}$$

where $t_{switch}$ and $t_{idle}$ are the amount of time when $p_k$ does task switching respectively is idle.

In an environment which allows task migration and dynamic scheduling policies the task set $T_k$ executed on $p_k$ and the slices $S_i$ are execution dependent so $E(p_k)$ cannot be accurately computed. In order to have an exact analytical model to reason about the overall throughput when having a certain cache partitioning a static assigning of tasks to the processors is required. In this way, tasks assigned to the same processor are executed overall sequentially, independent of the scheduling decisions for that processor and the processor's overall execution time will be:

$$E(p_k, T_k) = \sum_{t_i \in T_k} e(t_i, p_k, c(t_i)) + t_{switch} + t_{idle}$$

where $e(t_i, p_k, c(t_i))$ is the execution time of task $t_i$ on processor $p_k$.

To optimize the throughput, the task to processor assignment and the cache allocation should be such that $max(E(p_k, T_k))$ is minimized.

The power consumed by the system and the execution time are other possible optimization criteria for the multimedia systems. The consumed power depends by the time and the memory traffic that the system needs to complete all its tasks. Optimizing the overall execution time (respectively the number of misses) gives the most power consumptions reduction:

$$min\left(\sum_{t_i \in T} e(t_i, c(t_i))\right)$$

The method is applicable for systems with statically scheduling and allocation of tasks.

### 3.2. Proposed optimization method

Because in our experimental system task migration and dynamic scheduling are allowed, in the proposed method for finding the cache partitioning ratio the throughput and power consumed are improved by optimizing the overall application's number of L2 misses (so execution time).

The problem of minimization of the total number of cache misses is formulated as a (Mixed) Integer Linear problem. Let $m(t_i, c(t_i))$ be the number of misses of task $t_i$ with $c(t_i)$ allocated cache sets. The objective is to find $c(t_i)$ values for every task $t_i$ such that the overall number of cache misses is minimized:

$$min\left(\sum_{t_i \in T} m(t_i, c(t_i))\right)$$

We denote the set of valid cache sizes with $\{\sigma_k\}_{k=0,1,...,K-1}$ and $c(t_i) \in \{\sigma_k\}$. Due to implementation reasons $\sigma_k$ can be, for example, limited to powers of two. The number of misses of task $t_i$ with $\sigma_k$ cache sets ($m_i^k = m(t_i, \sigma_k)$) can be obtained by simulation or program analysis. In our model we use and average $\widetilde{m_i^k}$ over the $m_i^k$ obtained out of different simulation of task $t_i$ having $\sigma_k$ cache. We use a set $\{x_i^k\}_{(i=0,1...N-1,\ k=0,1,...,K-1)}$ of 0/1 variables that specify if the allocated cache for task $t_i$ is $\sigma_k$ or not. A task can have only one cache partition allocated, thus $\sum_{k=0}^{K-1} x_i^k = 1$.

As a result the ILP solver produces a $\{x_i^k\}$ assignment such that the overall number of misses is minimized. Using $\{x_i^k\}$ and $\widetilde{m_i^k}$ the number of misses for task $t_i$ can be written as:

$$m(t_i, c(t_i)) = \sum_{k=0}^{K-1} x_i^k \cdot \widetilde{m_i^k}$$

where the allocated cache size for every task $t_i$ is:

$$c(t_i) = \sum_{k=0}^{K-1} x_i^k \cdot \sigma_k$$

The sum of the cache sizes allocated to all tasks and communication buffers should be in the limits of the available cache size $C$. The size of the cache assigned to the communication buffer $b_l$ is $c(b_l) \in \{\sigma_k\}$ and this results in the following constraint:

$$\sum_{i=0}^{N-1} c(t_i) + \sum_{l=0}^{M-1} c(b_l) \leq C$$

## 4. Experimental framework

### 4.1. Application Model

To describe the applications we use Y-chart Applications Programmers Interface (YAPI) [2]. The model of computation in YAPI is based on Kahn Process Networks. Such a network consists of parallel tasks that communicate through (theoretically unbounded) FIFOs. Parallelism and communication are explicit and synchronization is done implicitly at read from empty FIFO (and write in full FIFO in the practical case where FIFOs are bounded). In practical video YAPI applications the "memory-active" entities of the system are tasks, FIFOs and frame buffers, so the cache is exclusively split among them. The application model is not restricted to YAPI descriptions and can be used with any multitasking programming model with shared memory primitives, e.g., POSIX threads.

The FIFOs access predictability is achieved by allocating them cache of the same size as the FIFO size.

For the frame buffers the production and consumption is done actually sequential (this is intrinsic for the YAPI application). The frames used for the prediction are accessed only after they are completely produced. This means that by allocating an exclusive cache partition for every frame buffer the compositionality of the system is preserved because the access to this buffer is sequential.

If all the tasks share the heap another factor that influences compositionality is dynamic memory allocation. Depending on which data was previously allocated (so depending on task scheduling and mapping) a tasks' data structure will have different address, resulting in different inside task misses. In the experiments in Section 5 we assume that the memory allocation is done during the initialization period and the overall allocation order is always the same.

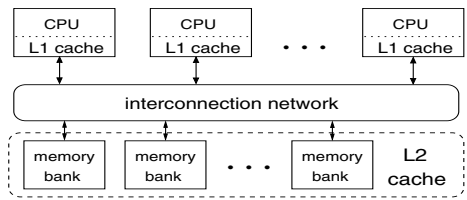

**Figure 1. CAKE architecture - inside-tile view**

### 4.2. Hardware platform and operating system

In the experiments a practical instance of the CAKE multiprocessor architecture is used [9]. The CAKE platform consists of a homogeneous network of computing tiles on a chip. Each tile contains CPUs (Trimedia and/or MIPS cores), a router (for out of tile communication) and memory banks (Figure 1). The processors are connected to memory by a fast, high-bandwidth snooping interconnection network. The on-tile memory is actually used as a L2 cache, shared between tasks, facilitating a fast access to the main memory which is outside the chip. The addressing space is linear.

Allocating sets of the L2 cache is implemented by changing the conventional index part of an address to a new index. In order to implement this, the cache has to be able to relate memory accesses to tasks and communication buffers. Ideally, each access should be labelled with a *task id* or *buffer id*. In this way the cache can translate the address according to a table indexed by this *id*. In the CAKE platform, the *task id* is kept in a register and can therefore be used directly.

There are several ways to obtain an *id* for communication buffers. A *buffer id* register could be used. This would imply additional requirements for the compiler, which then need to keep that register up to date. Alternatively, a part of the address could be used to encode the *buffer id*. This reduces the usable address space and also requires adaptation of the compiler for handling shared static data structures. Nevertheless, for dynamic memory allocation the partitioning can be implemented relatively straightforward by providing a dedicated malloc for shared buffers. A third approach is to keep a table with intervals of shared memory. This table needs to be loaded by the operating system. Then for every access the cache can lookup if the address has an associated *buffer id*.

The third alternative is chosen as for our experiments we are mainly interested in the system level aspects (e.g., inducing the compositionality property, implication in number of misses) and we are not concerned for the implementation's performance issues The third approach more generic than the others because any address range can be placed in any place in the cache. This easily allows for other experiments, like for example separating tasks' instructions, static initialized variables (data) and static uninitialized variables (bss) in the cache or sharing some cache partitions.

| tasks | FrontEnd1 | IDCT1 | Raster1 | BackEnd1 |
|---|---|---|---|---|
| alloc. L2 sets | 4 | 1 | 32 | 16 |
| tasks | FrontEnd2 | IDCT2 | Raster2 | BackEnd2 |
| alloc. L2 sets | 4 | 1 | 16 | 16 |
| tasks | Fr. canny | LowPass | HorizSobel | VertSobel |
| alloc. L2 sets | 4 | 16 | 8 | 16 |
| tasks | HorizNMS | VertNMS | MaxTreshold | |
| alloc. L2 sets | 8 | 8 | 4 | |
| data | appl_data | appl_bss | rt_data | rt_bss |
| alloc. L2 sets | 2 | 2 | 4 | 4 |

**Table 1. L2 allocated sets for 2 jpegs & canny**

| tasks | input | vld | hdr | isiq | memMan |
|---|---|---|---|---|---|
| alloc. L2 sets | 2 | 4 | 16 | 8 | 1 |
| tasks | idct | add | decMV | predict | predictRD |
| alloc. L2 sets | 4 | 4 | 8 | 16 | 2 |
| tasks | writeMB | store | output | | |
| alloc. L2 sets | 8 | 2 | 1 | | |
| data | appl_data | appl_bss | rt_data | rt_bss | |
| alloc. L2 sets | 4 | 1 | 8 | 1 | |

**Table 2. L2 allocated sets to tasks for mpeg2**

We have adapted the operating system, such that it manages the necessary translation tables for the cache. For this, it offers primitives of cache allocation for tasks and for shared memory.

## 5. Experimental results

We evaluate the proposed technique using as workload two applications. The first application (15 tasks) consists of two jpeg decoders [1] working on different pictures formats and one line based canny edge detection algorithm. The second application (13 tasks) is a mpeg2 video decoder [11]. In both examples there is a run-time operating system that has an exclusive cache part allocated such that it does not interfere with the application's tasks.

The instance of the CAKE platform used has four Trimedia processors, 512KB, 4 ways associative L2. On the experimental platform the application and run time system static allocated data (data and bss) is shared between tasks so in order to have predictable access, with the same consideration as for communication buffers (Section 3), exclusive cache partitions are allocated for them as well. In the last row of Tables 1 and 2 (corresponding to the two examples) the allocated cache sizes are presented.

We performed simulations in order to determine the number of misses depending of the allocated cache size (for $\sigma_k = 2_\alpha$) and we applied the method in Section 3.2. The obtained partitioning ratios are presented for the two examples in Table 1 and Table 2. To give a measure of the inter-task cache interaction in Figure 2 the number of misses for the shared cache case and partitioned cache case are presented for all tasks and communication buffers. For both examples we can observe that with only few sets of exclusive cache assigned to static allocated data a major improvement in performance is obtained.

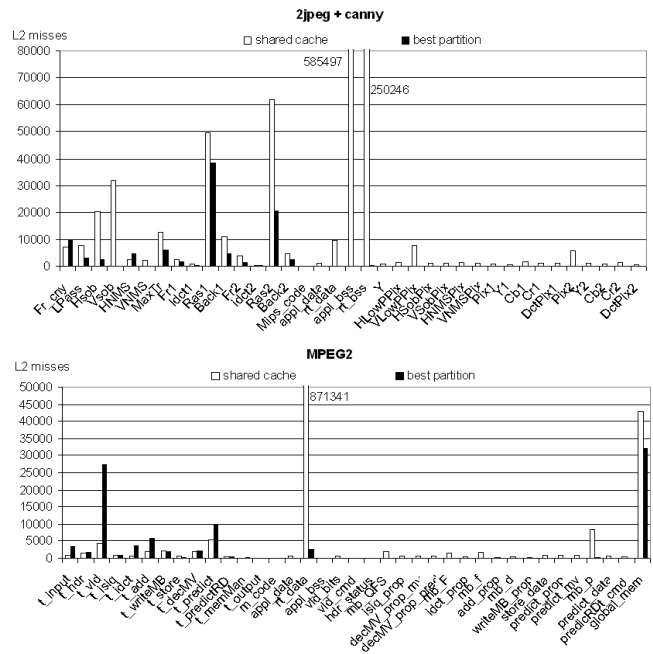

**Figure 2. Shared vs. best partitioned cache for every task and communication buffer**

With respect to performance cache partitioning has two effects: on one hand the misses due to inter-task conflicts are alleviated but on the other hand some tasks do not benefit from all the cache space, increasing their number of misses. The balance between the two effects can be seen in Figure 2. The results also indicate that for the first application the L2 miss rate is improved from 9.46% to 2.21% and as a result the number of cycles per instruction (CPI) of every processor is reduced with approximate 20% (from 1.4 cpi to 1.1 cpi). For the mpeg2 application the L2 miss rate is improved from 5.1% to 0.8% and as a result the number of CPI of every processor is reduced with approximate 4% (from 1.7-1.8 cpi to 1.6-1.7 cpi). For this application the reduction in CPI is not so large because the used mpeg2 implementation was not optimized for Trimedia so it is more L1 and processor bounded than L2 bounded. The mpeg2 was also simulated with 1MB of shared L2 cache and for that the L2 miss rate was 0.6% with a corresponding 1.7 cpi for every processor.

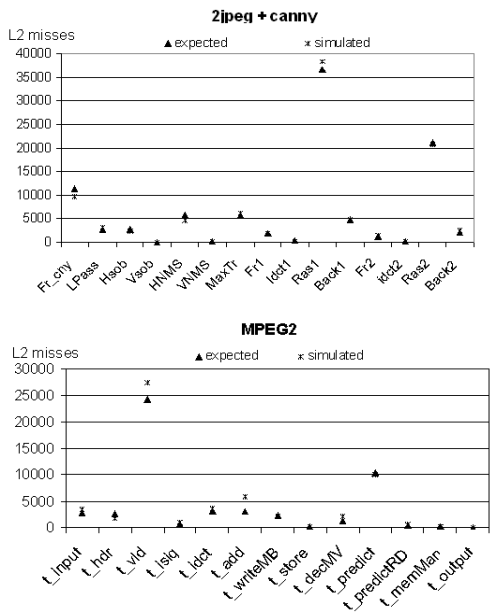

**Figure 3. Expected-simulated performance comparison for every task**

The graphs in Figure 3 presents the difference between the number of misses for every task $(\widetilde{m_i^k})$ expected using the model from Section 3.2 and the ones obtained by simulation of the determined best cache partitioning ratio for the application. These differences give a measure of compositionality and reflect the variation in number of L2 misses due to the neglected effects like task switching, L1 and bus contention. For both examples the largest difference for a task between the expected and simulated number of misses relative to the overall simulated number of misses is 2%. The results suggest that system behaves as expected and the compositionality property is induced.

## 6. Conclusions

For multimedia applications when running on embedded multiprocessor systems we tackle the problem of non compositionality due to inter task conflicts in the shared cache. In the proposed approach compositionality is induced by partitioning the unified level of cache shared between processors using a scheme that exclusively allocates sets of cache to the tasks and inter-task communication buffers. Different optimization criteria, statically prioritizing tasks and guaranteeing performance are possible by tuning the cache partitioning ratio. For homogeneous multiprocessors the problems of finding the cache partitioning ratio for optimizing the throughput and the power are analytically formulated and the influence of task to processor assignment is discussed. A practical approximation of these formulations is presented and it is validated using two applications: two jpegs decoders and a canny edge detection running in parallel and a parallel mpeg2 decoder. Applying the proposed method to a CAKE instance with 4 Trimedia processors and 512KB L2 cache compositionality was achieved. The largest difference for a task between the expected and simulated number of misses relative to the overall simulated number of misses is 2%. In terms of performance, compared with the shared cache case there were 5 times less L2 cache misses resulting in 20% reduction in cycles per instruction for the first example and 6.5 times less L2 cache misses resulting in a 4% cpi reduction for the second example.